\documentclass[aps,prl,onecolumn,groupedaddress]{revtex4}
\usepackage{graphicx}
\usepackage{dcolumn}
\usepackage{bm}
\usepackage[english]{babel}
\usepackage{amsmath}
\usepackage{amssymb}
\usepackage{amsfonts}
\usepackage{graphics}
\usepackage[latin1]{inputenc}
\usepackage[T1]{fontenc}
\usepackage[colorlinks=true]{hyperref}

\begin{document}

\title{Ground states of anisotropic antiferromagnets with
single ion  and cubic anisotropy}


\author{T.-C.~Dinh}
\email[]{dinh_thanh_chung@yahoo.de}
\affiliation{Institut f\"ur Theoretische Physik, Johannes Kepler
Universit\"at Linz, 4040 Linz, Austria }
\author{R.~Folk}
\email[]{reinhard.folk@jku.at}
\affiliation{Institut f\"ur Theoretische Physik, Johannes Kepler
Universit\"at Linz, 4040 Linz, Austria }

\date{\today}
\begin{abstract}
 Anisotropic antiferromagnets in an external magnetic field show a rich variety
of different ground states meeting in transition lines and multicritical points. 
We study the dependence of the ground states of these systems in the three dimensional space
on physical parameters as exchange, single ion and cubic anisotropy. 

One identifies four different ground states: the paramagnetic (PM), the antiferromagnetic (AF),
 the spin flop (SF)  and the
biconical (BC) ground state. In the case of absence of a cubic anisotropy the transition lines separating the
different ground states can be calculated analytically, otherwise they have to be calculated numerically.
We also considered the behavior of the staggered magnetization which characterizes the different ground states.
From its behavior the order of the transition from one state to the other is determined.

But also the order of the transition changes along the transition 
lines when including the cubic anisotropy,  especially at the reeentrant region where
a transition from SF to BC and back to SF by increasing the external field $H$ occurs.
Multicritical points are founded which are assumed to be tricritical or critical
endpoints.

The results obtained may be relevant for other systems since the antiferromagnetic model can be mapped to a
lattice gas model where the biconical ground state is interpreted as supersolid phase. Recent
renormalization group calculations show that such a phase would indicate the existence of a
tetracritical point.
\end{abstract}
\maketitle


\section{Introduction}

Anisotropic systems in an external field not coupled to the order parameters (OP) characterizing
the condensed phases show interesting phase diagrams depending on the values of the anisotropy and the
external field. One example is the anisotropic antiferromagnet in an external magnetic field \cite{shapira-multicritical}.
Other ones  are crystal systems undergoing a displacive phase transition under external pressure
\cite{BruceAharony75}. The possible variety of phases at finite temperature is also determined by the
topology of the ground state phase  diagram since the phases corresponding to the different ground states
extend to finite temperatures. However at finite temperature new phases not present at $T=0$  may
arise \cite{LiuFisher73}. Due to the variety of possible ground states or phases multicritical points
of different kind are possible at finite $T$. Using the magnetic language this systems show
antiferromagnetic (AF), spinflop (SF), biconical (BC) and paramagnetic (PM) ground states and phases.
The transition lines between these phases may meet in one point: (i) the bicritical point, where three
phases coexist (AF, SP, and PM) and (ii) the tetracritical point, where four phases (AF, BC, SP, and PM)
coexist. Going beyond mean field theory \cite{LiuFisher73} the nature of the multicritical points has
been studied within renormalization group theory \cite{nekofi74} (in a one loop order calculation).
These results have been questioned by higher loop calculations \cite{Prudnikov98, capevi03}. Recently
both using simulation methods \cite{Selkeetal08, Holtschneider} and renormalization group theory \cite{fohomo08,FHM08}
the nature of the multicritical point have been reconsidered. 

It is well known that there is a correspondence between such magnetic model and the quantum lattice
gas \cite{fisher67} which allows to transpose the results obtained for the magnetic system to systems
like He$^4$ with superfluid and supersolid phases \cite{peters09}. The question of the existence of
such phases is of general interest.

Symmetry consideration are essential for the existence and the nature of ground states and phases.
This is also the case for the systems treated here. It has been found that interaction terms of
cubic symmetry has strong influence on the phase diagram \cite{Selke08}. Here we present results
at $T=0$ which clarifies the order of the transition lines between the different ground states.
Moreover we extend the models studied so far by combining several types of anisotropy. 

\begin{table}
\begin{tabular}{|c|c|c|c|c|c|}
\hline
Phase           & $N_\|$ & $N_\perp$       & $M_\|$    &$M_\perp$   & Phase of QLG-model    \\ 
\hline
AF           & $\neq 0$ & $0$       & $0$    &$0$   &   solid    \\ 
SF           & $0$      & $\neq 0$  & $\neq 0$ & $0$ & superfluid     \\ 
BC           & $\neq 0$  & $\neq 0$ & $\neq 0$ &  $\neq 0$ & super solid   \\ 
PM           & $0$       &  $0$     & $\neq 0$ &  $0$ &  fluid\\  
\hline
\end{tabular}
\caption{Correspondence between the phases of the classical magnetic system and those of 
the quantum lattice gas (QLG).
Also shown are the order parameters, the alternating magnetization in the fied direction $N_\|$ and perpendicular to it $N_\perp$, and 
 the  components of the magnetization $M_\|$ and $M_\perp$}
\label{compare}
\end{table}

\section{Single ion anisotropy} 

Let us first consider the well studied case of an antiferromagnet with the Hamiltonian
${\cal H}$
 \begin{equation} \label{Hsingl}  
 \mathcal{H}=J\sum_{\langle ij \rangle}^{N_{A},N_{B}}\left( 
\Delta \left(S_{ix}S_{jx} + S_{iy}S_{jy} \right) 
+S_{iz}S_{jz} \right)
-H\left( \sum_{i}^{N_{A}}S_{iz} + \sum_{j}^{N_{B}}S_{jz} \right)
+D\left(\sum_{i}^{N_{A}}S^{2}_{iz}
   + \sum_{j}^{N_{B}}S^{2}_{jz}\right)
\end{equation}
It describes $N_A$ and $N_B$ interacting classical spins $\vec{S}_i$ on two sublattices $A$ and $B$,
where $N_A=N_B$. The anisotropy of the positive exchange interaction $J$ is characterised by the parameter
$\Delta$ with $\Delta=0$  for a magnet where only the $z$-components of the three dimensional spins interact
 and $\Delta=1$ for an isotropic  magnet. The spins are subjected to an external magnetic field $H$
and  a single-ion anisotropy $D$.

\begin{figure} \hspace{-1.6cm}
  \includegraphics[height=.25\textheight]{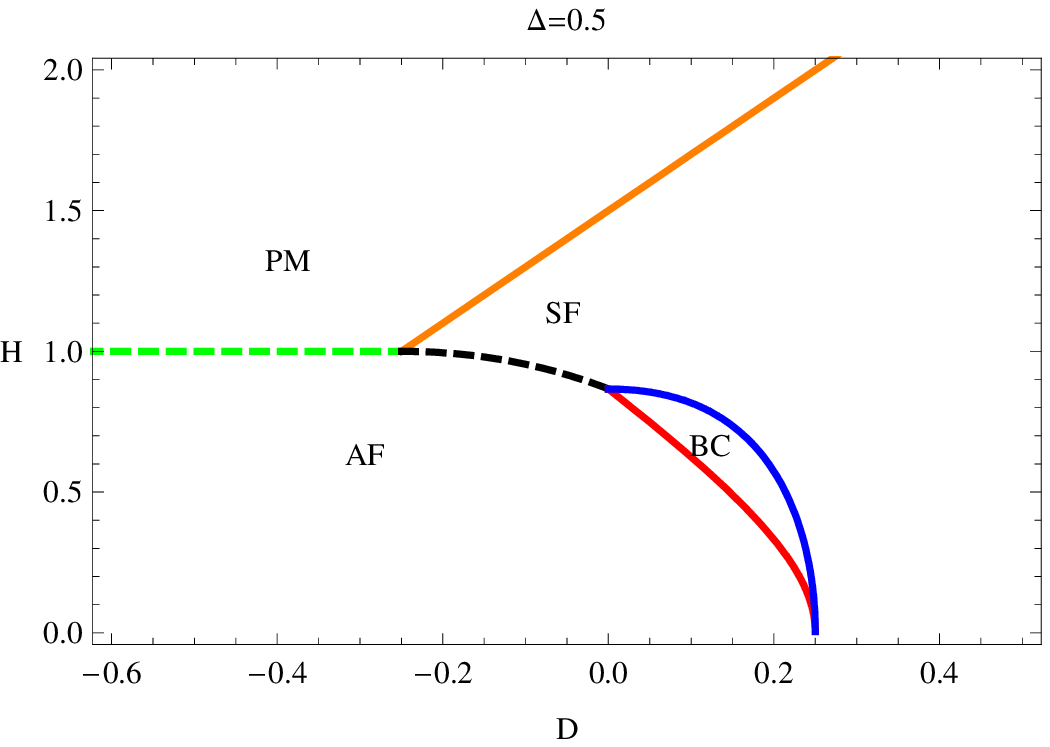}\hspace{-0.2cm}
\includegraphics[height=.25\textheight]{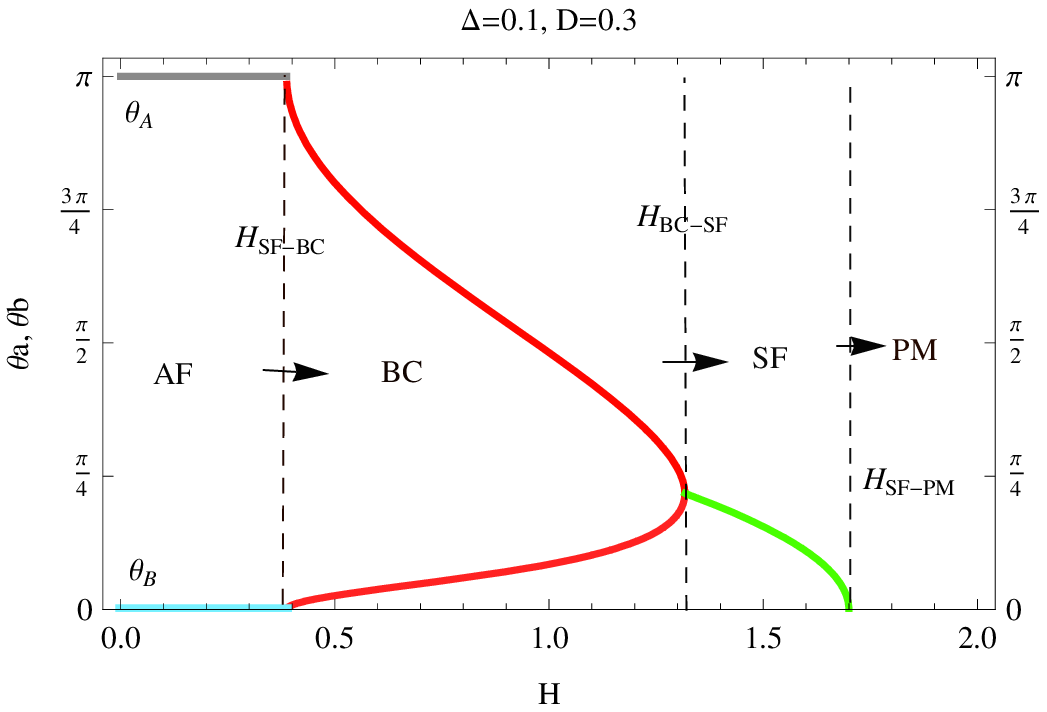} 
  \caption{(online color)Left: The phase diagram for the XXZ-Model including the single-ion-anisotropy. Right: The 
order parameter of the four ground states configuration can be calculated analytically.
\label{ion}}
\end{figure}

The case where beside the exchange anisotropy only a single-ion anisotropy is present can
be solved analytically \cite{matsuda70} at $T=0$. A BC ground state exists for a certain region of positive $D$
and external magnetic field $H$ for  $\Delta=0$. This region becomes  smaller for increasing $\Delta$ and
diminishes for an isotropic antiferromagnet with $\Delta=1$. The transition lines between
the AF to the PM  and between the AF and the SF ground state are of
first order. The other transitions are of second order.
Figure \ref{ion} shows 
the whole ground state diagram with the parameters chosen so that a BC ground state appears.
Here we divide the parameters $H,D,F$ by the number
of next neighbors ( $z=6$ for simple cubic system), so in order to compare with the results in  \cite{Selke08,Bannasch08}
one has to multiply all the three parameters in this paper with $6$.
Also the OP for the complicated BC ground state can be calculated exactly which reads:
\begin{eqnarray}
\cos\theta_{A,B}(H,D,\Delta)=
\frac{H \pm\sqrt{\frac{\sqrt{(1-2 D)^2-\Delta ^2} \left((2
   D+1)^2+H^2-\Delta ^2\right)+2 H \left(4 D^2+\Delta
   ^2-1\right)}{\sqrt{(1-2 D)^2-\Delta ^2}}}-\sqrt{(1-2
   D)^2-\Delta ^2}}{4 D}  
\end{eqnarray}

An extension to finite temperature within mean field theory has been discussed in \cite{LiuFisher73}. As
already explained  in \cite{fisher67} and used in \cite{matsuda70} a {\it quantal} lattice gas can be mapped
on  an anisotropic antiferromagnet  with a Hamiltonian of form  (\ref{Hsingl}). The correspondence of the different
ground states is given in Tab. \ref{compare}.

In the $H-T$ ground state  diagram the ground states at $T=0$ extend as phases to higher temperature. Since for high 
enough temperature
and large enough external magnetic field the PM phase should be the stable one, all other phases are confined
to finite regions 
in the phase diagram. Thus it happens that the phase transition lines cross each other in multicritical 
points. This might be a bicritical point when the AF, the SF and the PM
phase meet,
or a tetracritical point where the AF, the SF, the BC and the PM phase
meet.
It should be noted that this is even possible if the BC phase does not extend to $T=0$
\cite{LiuFisher73}.

\section{Cubic anisotropy}
\begin{figure} \hspace{-1.5cm}
  \includegraphics[height=.25\textheight]{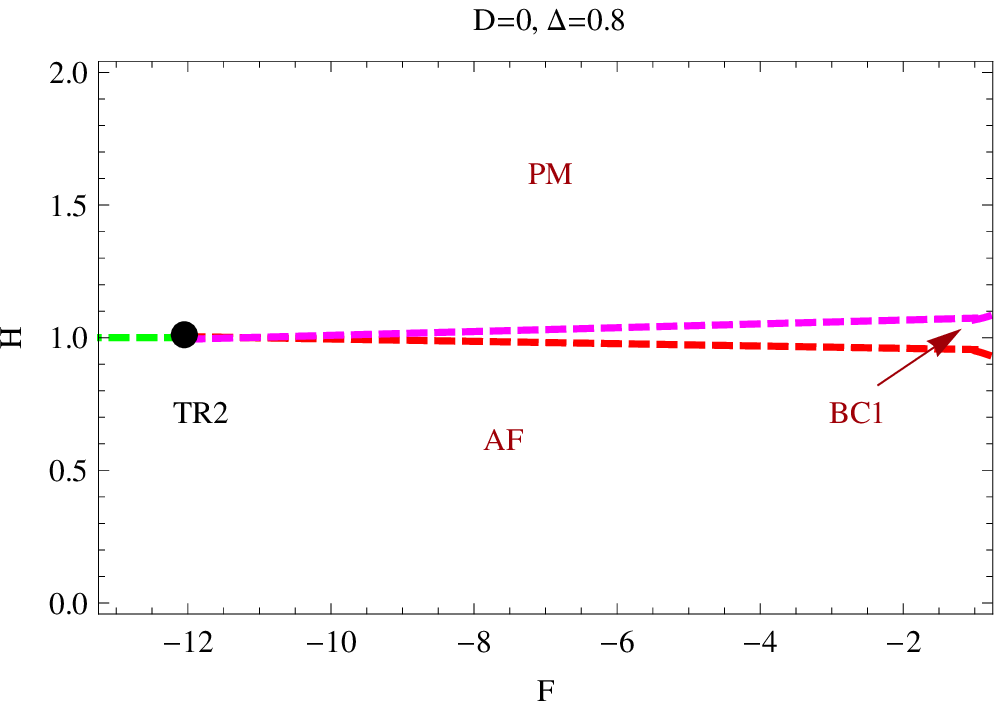}\hspace{-1cm}
\includegraphics[height=.25\textheight]{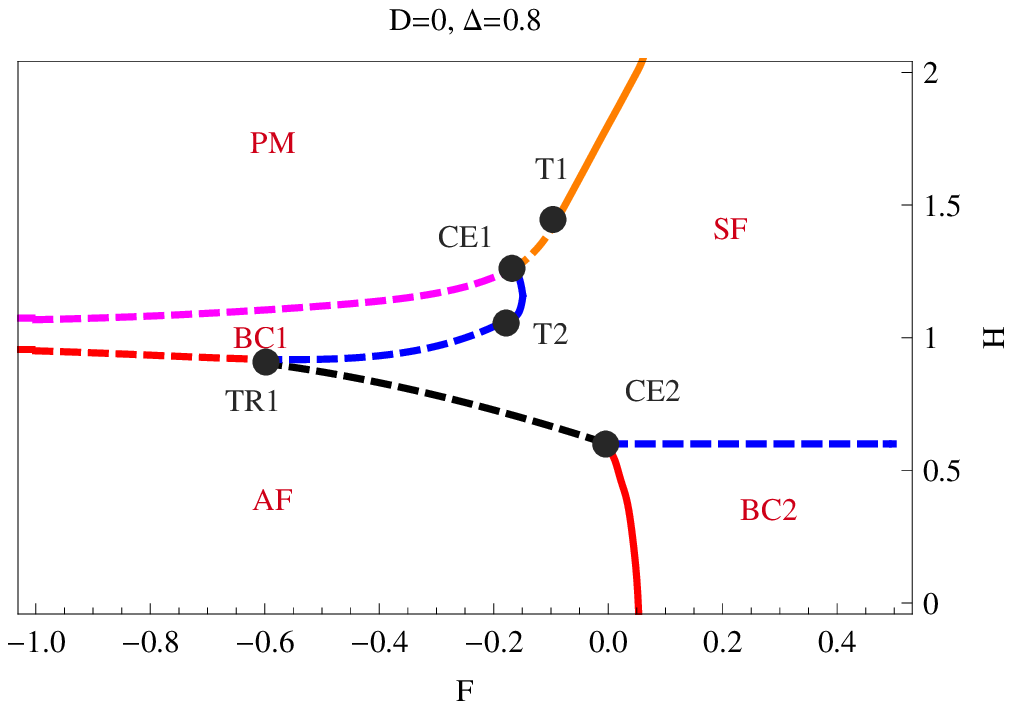} 
  \caption{(online color)Ground states minimizing the energy, Eq. (\ref{Hcubic}), in the magnetic field $H$ and
cubic anisotropy $F$ plane at an exchange anisotropy $\Delta=0.8$ and with zero single ion anisotropy $D$.
The ground states are denoted as follows: AF antiferromagnetic, PM paramagnetic, SF spin flop ; BC1,2 biconical
oriented along the diagonal or the axis in the $x-y$-plane. Shown are special points where either
more than two ground states  meet or where the order of the transition line changes from first (dashed line)
to second order (solid line). These are the triple points TR1,2, the tricritical points T1,2 and the
critical end points CE1,2.
\label{cub1}}
\end{figure}

Uniaxial antiferromagnets may also have cubic anisotropy. Here we consider an anisotropic antiferromagnet
with a
cubic anisotropy $F$ instead of the single-ion anisotropy $D$. For such a system the Hamiltonian reads
\begin{eqnarray} \label{Hcubic}  
 \mathcal{H}=J\sum_{\langle ij \rangle}^{N_{A},N_{B}}\left( 
\Delta \left(S_{ix}S_{jx} + S_{iy}S_{jy} \right) 
+S_{iz}S_{jz} \right)
-H\left( \sum_{i}^{N_{A}}S_{iz} + \sum_{j}^{N_{B}}S_{jz} \right) \nonumber \\
+F\left(\sum_{i}^{N_{A}}\left(S^{4}_{ix}+S^{4}_{iy}+S^{4}_{iz}\right) 
   + \sum_{j}^{N_{B}}\left(S^{4}_{jx}+S^{4}_{jy}+S^{4}_{jz}\right)\right) \, . 
\end{eqnarray} 
The effect of such a type of the cubic anisotropy has been studied recently by \cite{Selke08,Bannasch08}. It 
has been shown that two types of BC  ground states are possible (BC1 for $F<0$ and BC2 for $F>0$), depending
whether the perpendicular component of the alternating magnetization lies in the  diagonals or along the
axes in the $xy$-plane. Moreover it has been
found that most of the transition lines between the  ground states  become first order. There was also
found a reentrant region for the SF ground state becoming unstable for  increasing the external magnetic
field with respect to the BC state, which for even larger field again becomes unstable with 
respect to the SF phase (see Fig. \ref{cub1}). 
 In this reentrant region the transition line has been
found to be of second order.

\begin{figure} \hspace{-1.5cm}
  \includegraphics[height=.25\textheight]{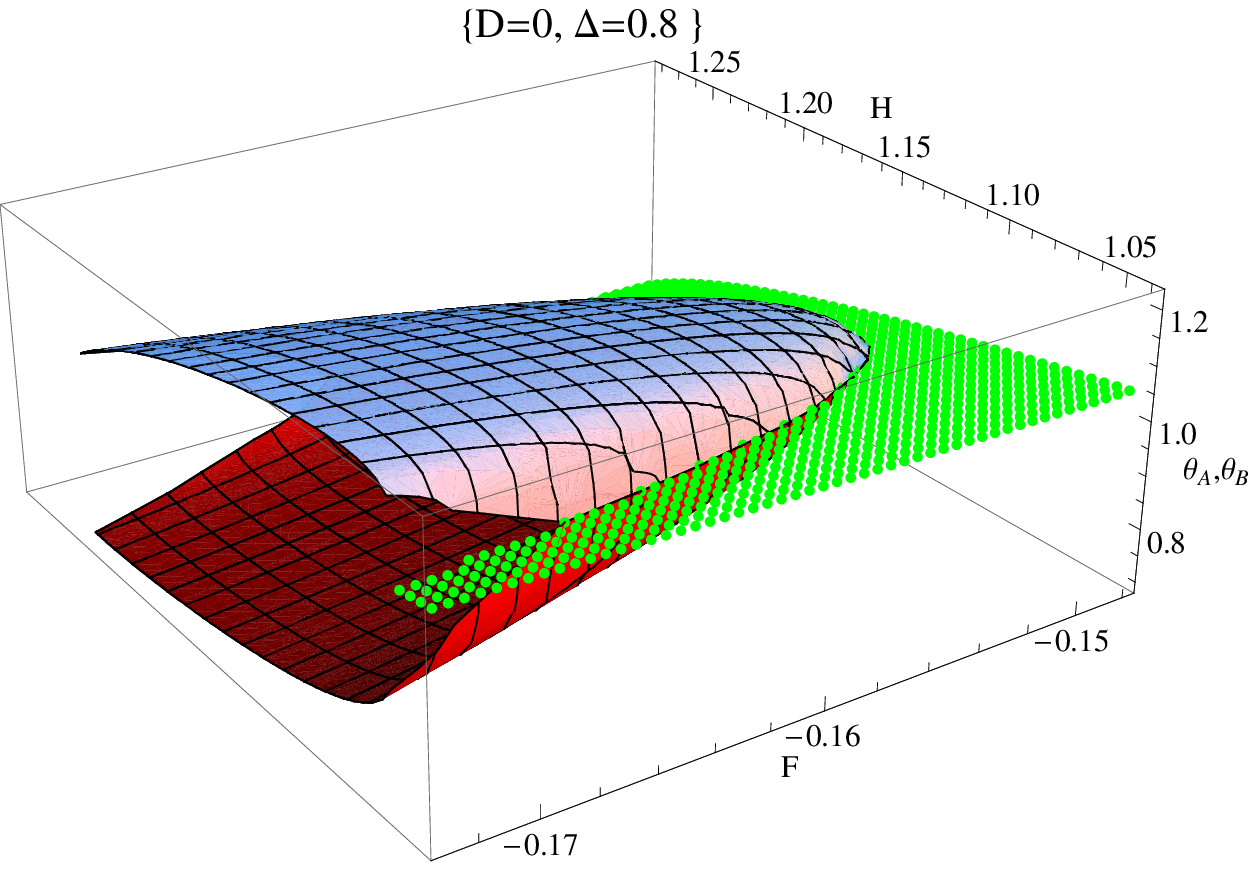}\hspace{-0.3cm}
\includegraphics[height=.25\textheight]{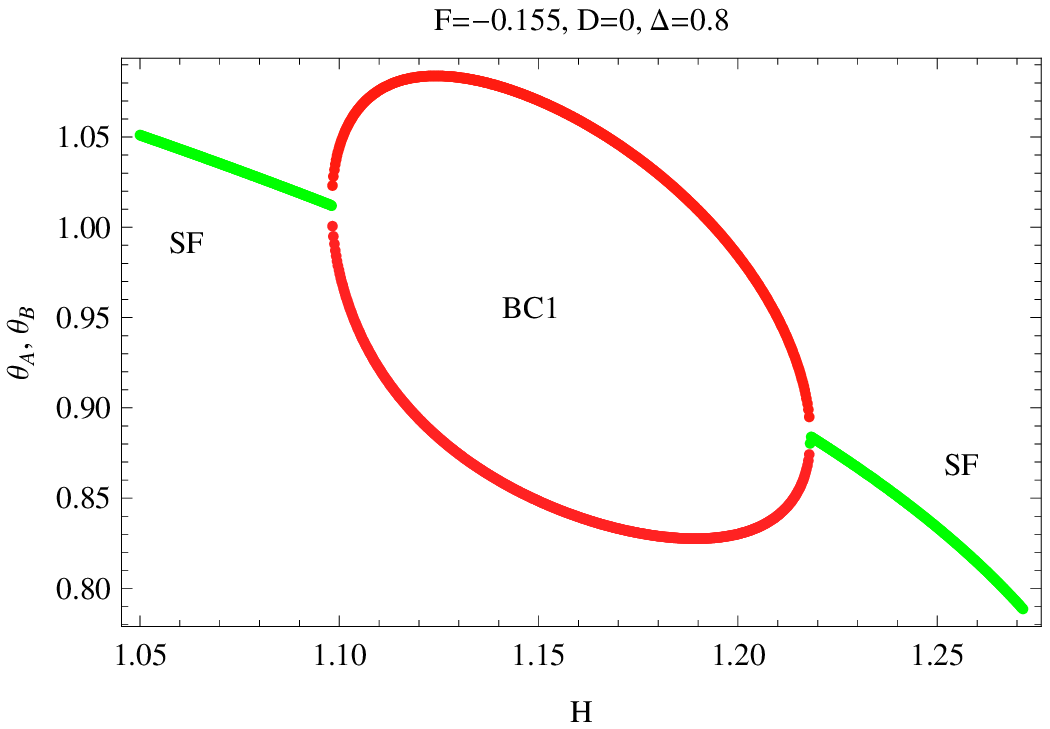} 
  \caption{(online color)A 3D and a 2D diagram showing the occurrence of the reentrant by increasing 
the magnetic field $H$
\label{ordcub}}
\end{figure}

We verify the occurrence of the reentrant region \cite{Selke08} by calculating the order parameter of 
the ground states which take place in the vicinity of 
this region (see Fig~\ref{ordcub}). The 2D figure shows the the SF-BC1 occurs
by increasing the applying field $H$. Continue to increasing the field further 
the SF reappears again.

\begin{figure}  \hspace{-1.5cm}
  \includegraphics[height=.22\textheight]{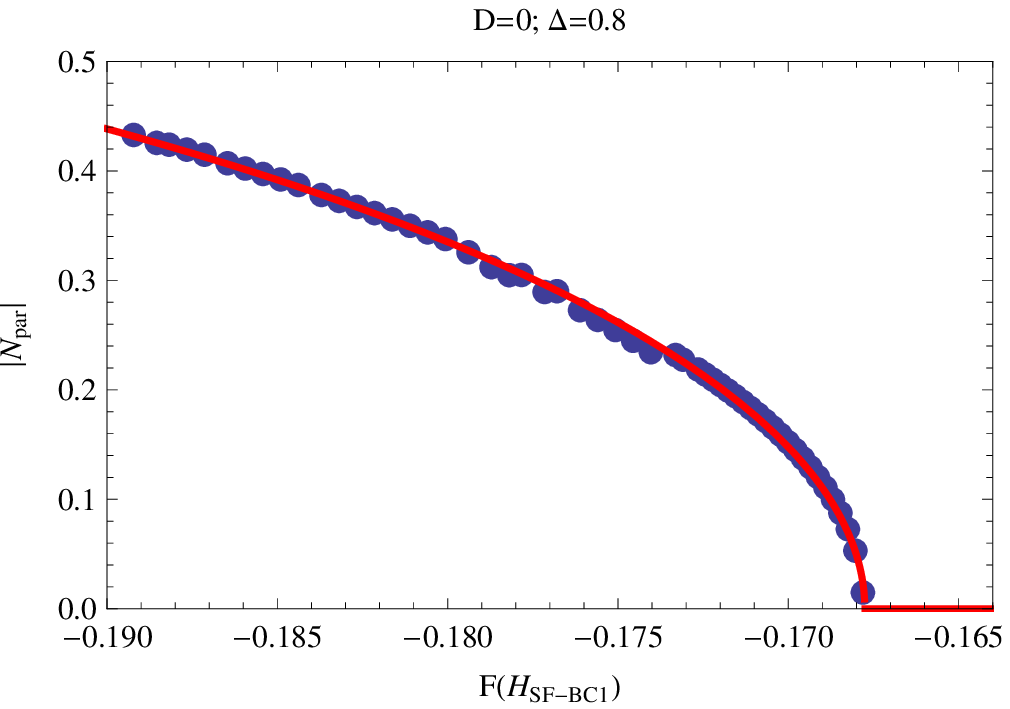}
   \includegraphics[height=.22\textheight]{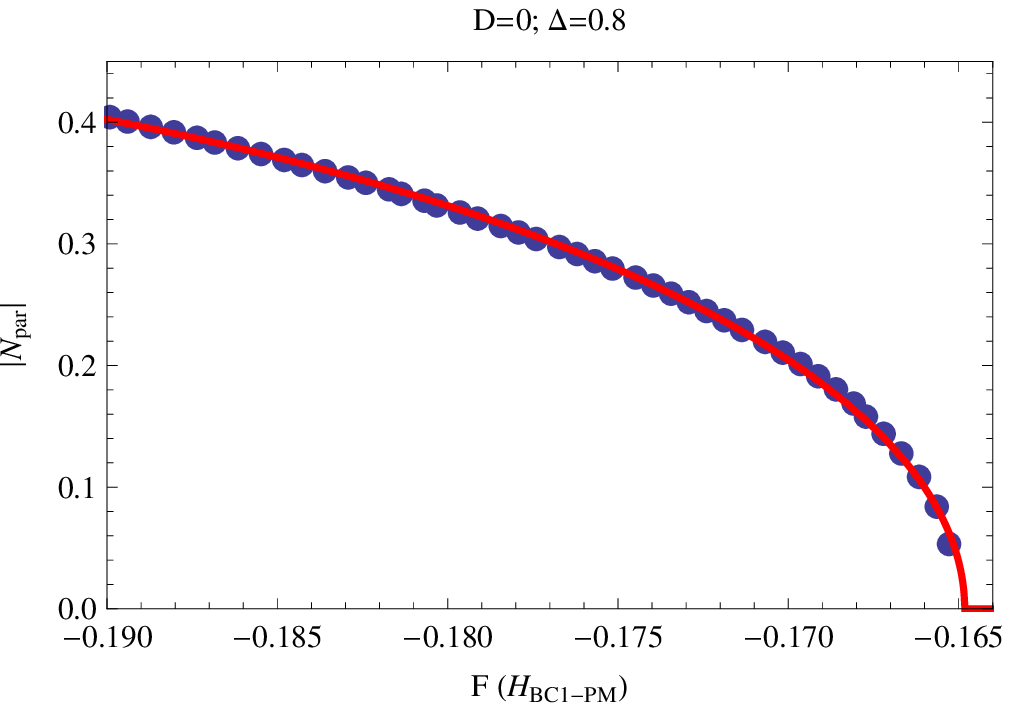}
  \caption{(online color)Left: Change of the parallel order parameter $|N_\||$ going along the first order transition lines between the BC1 and PM 
ground states. The dots are calculated values and the solid line a fit according to Eq (\ref{fit}). The zero value of the 
parallel OP is reached at the critical end point CE1. Right: Similar the change of the perpendicular order parameter $|N_\perp|$ going 
along the first order transition lines between the the SF and PM ground state. The zero value of the 
perpendicular OP is reached at the tricritical point T1.
\label{op}}
\end{figure}

In order to find out the transition points where a change in the order of the transition takes place
we have to calculate the nonzero order parameter component along the border lines for the case treated in
\cite{Selke08}. Along the 
transition line between the BC1 and the PM ground state the parallel staggered magnetization $N_\|$ 
goes to zero at the point where the
SF - PM and the BC1- SF transition lines meet (see Fig.\ref{cub1}). The behavior of the OP is compatible with a
power law with a correction term
\begin{equation} \label{fit}
|N_{\perp,\|}|=A\left|\frac{|F(H)-F^\star|}{F^\star}\right|^{0.5}\left(1+B\left|\frac{F(H)-F^\star}{F^\star}\right|\right)
\end{equation}
where $F^\star$ is the value of the cubic anisotropy at the special point, and $F(H)$  the value of the cubic
anisotropy along the
transition line. $A$, $B$ and $F^\star$ are fit parameters. This point is a critical end point
(CE1). From this critical endpoint the second order reentrant line starts. This line ends in a point where the AF, the BC1 
and the SF ground state meet.  This point turns out to be a triple point (TR1) since the second order character
of the reentrant line is changed to first order (at T2) before it reaches the triple point.
The value of $N_\|$ becomes nonzero on the reentrant line at T2 and this point might be a tricritical point
(see Fig. \ref{op}).

\begin{figure} \hspace{-1.5cm}
  \includegraphics[height=.22\textheight]{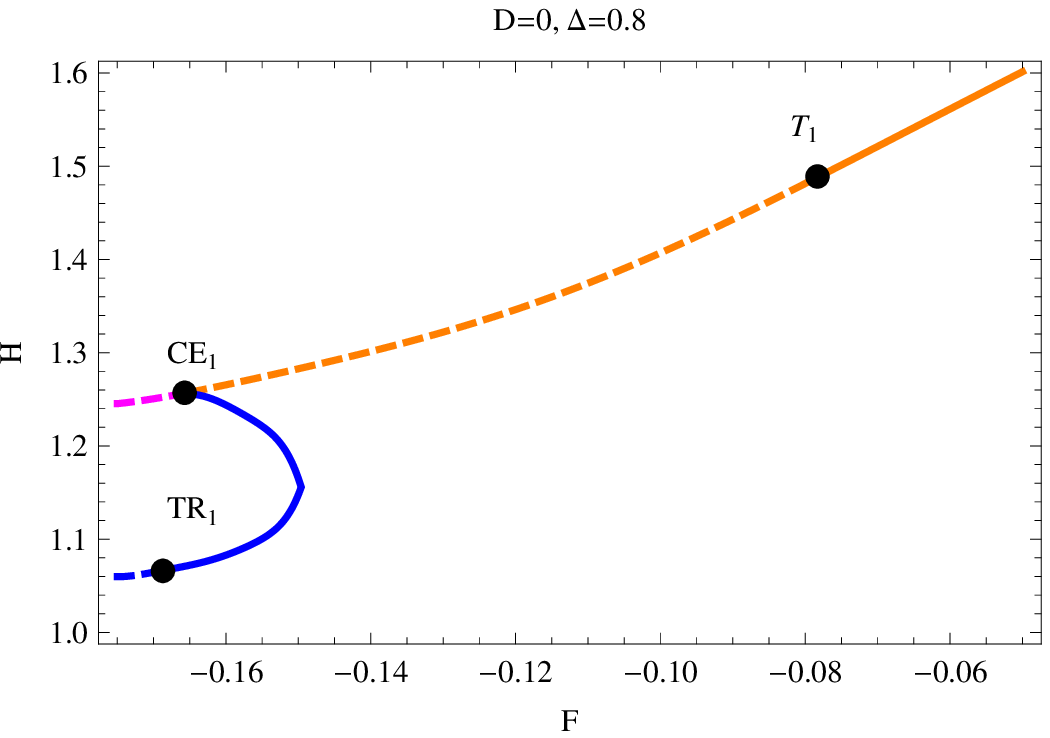}\hspace{-0.3cm}
\includegraphics[height=.22\textheight]{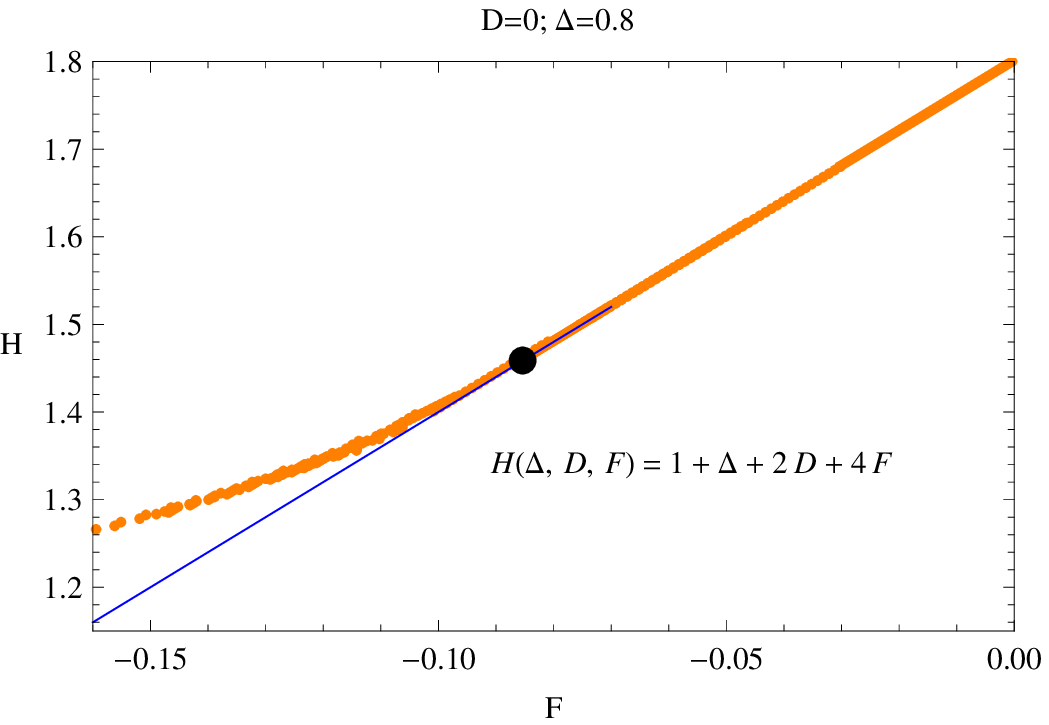} 
  \caption{(online color)Left: Detail from the right part of Fig. 1 on a larger scale around the reentrant region.
Solid lines are of second order, dashed lines of first order. The second order reentrant line between the biconical
and spin flop ground state starting at the tricritical  point TR1 ends in the critical end point CE1 on the
transition line to the paramagnetic ground state. On the transition line between the paramagnetic and spin flop
ground state one finds the tricritical point T1. 
Right: Comparison of the transition line between the SF and PM ground state calculated on the
assumption of a second order transition (straight line) and the transition line calculated numerically 
by minimizing the energy.
Indeed when the  transition curve deviates from the second order line the transition becomes first order.
\label{cub2}}
\end{figure}

We also find a change of order along the transitions line between the SF and the PM ground state.
At T1 the transition becomes second order for larger values of $F(H)$. This is just at
the point where  the transition line  begins to deviate from the analytically calculated
second order transition line (see Fig. \ref{op}).
                  
Conclusively we summarize all the critical points in a table 
(see Tab~\ref{tab:critpoint})
\begin{table}
\caption{(online color)\label{tab:critpoint}Summary of all the special points
discovered in the ground state diagram for $D=0$ and $\Delta=0.8$.
There exist tricritical points (T),  tripel points (TR) and 
the critical end points (CE).}
\begin{tabular}{ll|ll}
\hline \hline
Critical Points & Coordinates  ($F,H$)&Critical Points & Coordinates  ($F,H$)\\ \hline
T1&$(-0.087747,1.49)$&TR2&$(-12,1)$\\
T2&$(-0.167755 ,1)$&CE1&$(-0.164823,1.26)$\\
TR1&$(-0.6,1.6)$&CE2&$(0,0.6)$\\
\hline \hline
\end{tabular}
\end{table}
 
\section{Single ion and cubic anisotropy}
\begin{figure} \hspace{-1.5cm} 
  \includegraphics[height=.22\textheight]{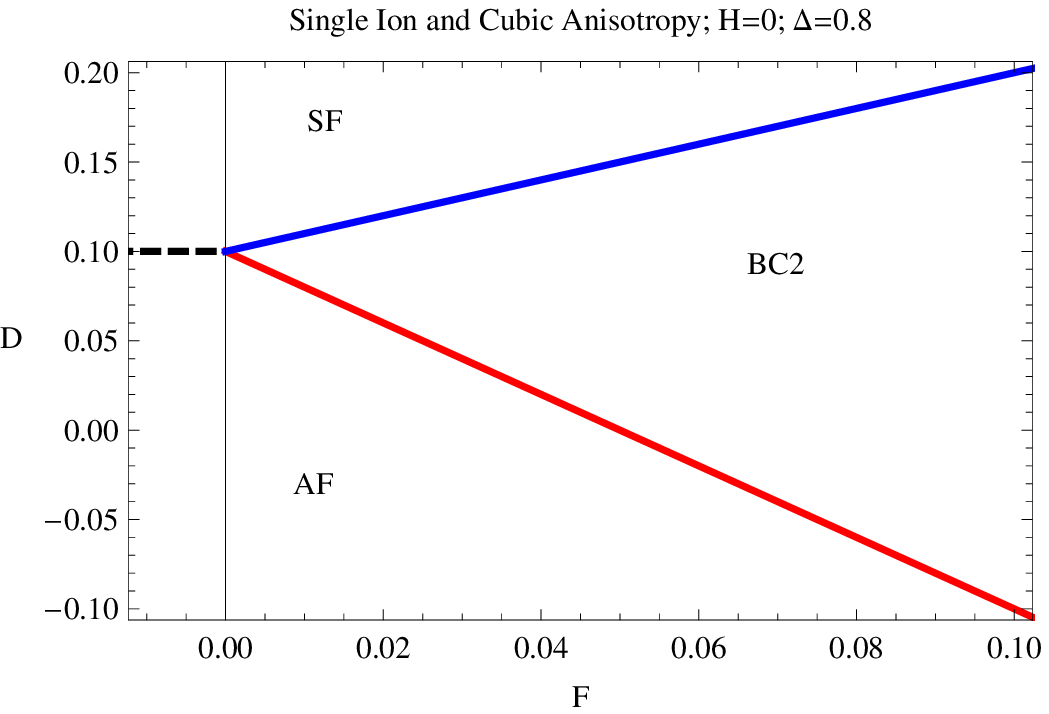}
\includegraphics[height=.25\textheight]{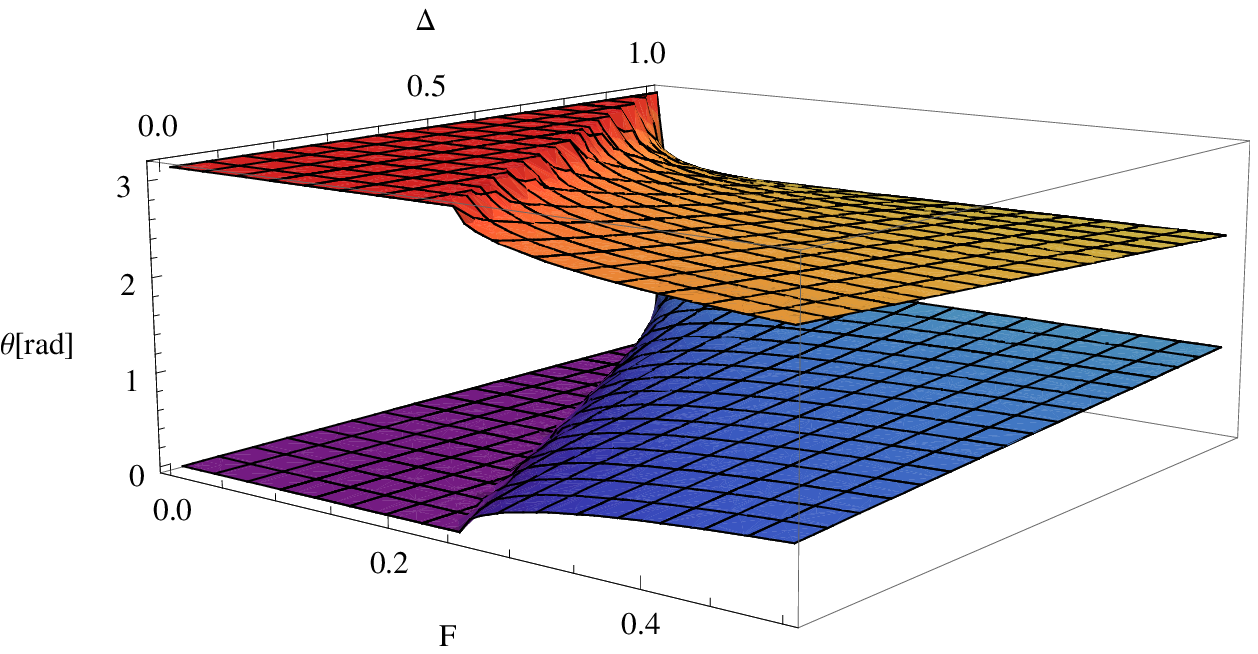} 
  \caption{(online color)Left: The analytically calculated ground state diagram for $H=0$. Right:
The continuous transition
from AF to BC2. The two polar angles, from which the OP can be calculated,  of the two different  sublattices  are
 shown \label{hfz0}}
\end{figure}

Let us now consider the general case when the exchange, single-ion and cubic anisotropy are present 
\begin{eqnarray} \label{all}  
 \mathcal{H}=J\sum_{\langle ij \rangle}^{N_{A},N_{B}}\left( 
\Delta \left(S_{ix}S_{jx} + S_{iy}S_{jy} \right) 
+S_{iz}S_{jz} \right)
-H\left( \sum_{i}^{N_{A}}S_{iz} + \sum_{j}^{N_{B}}S_{jz} \right) \nonumber \\
D\left(\sum_{i}^{N_{A}}S^{2}_{iz}
   + \sum_{j}^{N_{B}}S^{2}_{jz}\right)  
+F\left(\sum_{i}^{N_{A}}\left(S^{4}_{ix}+S^{4}_{iy}+S^{4}_{iz}\right) 
   + \sum_{j}^{N_{B}}\left(S^{4}_{jx}+S^{4}_{jy}+S^{4}_{jz}\right)\right) \, . 
\end{eqnarray} 
Then the planar ground state regions extend to three dimensional regions and  whole lines of
multicritical points appear. In Fig. \ref{all} we give as example the case where the single ion
anisotropy has been chosen nonzero ($\Delta=0.8$) and a sketch of the three dimensional ground state diagram. 
An analytic result for the transition line between the SF ground state and the PM ground state
can be  calculated assuming a second order transition, which of course is just correct in certain
section of the transition line (see Fig.~\ref{cub2}). The relation reads
\begin{equation}
H(\Delta, D, F)=2D+4F+\Delta+1
\end{equation}
Also in zero magnetic field ($H=0$) the stable ground states as function of the anisotropies 
can be calculated analytically.  The transition between the AF  and the BC2 ground state
starts at a positive value of the cubic anisotropy $F$
\begin{equation}
F(D,\Delta)=(1-2D-\Delta)/4.
\end{equation}
The transition occurs just only in the region $D<\frac{1-\Delta}{2}$
 and is continuous (see Fig \ref{hfz0}), where the OP parameter of 
the biconical ground state (BC2) can be given.
\begin{equation}
\theta_{A,B}=\arccos \left(\pm\frac{1+2F -2D - \Delta}{6 F}\right)
\end{equation}
This 2nd order transition continues to be valid also when applying when magnetic field.
Increasing the single ion isotropy ( $D>\frac{1-\Delta}{2}$ )
the Spin-Flop is preferred to the BC2 and
the transition takes place at 
\begin{eqnarray}
F(D,\Delta)=\frac{-1+2D+\Delta}{2}
\end{eqnarray}
If the cubic anisotropy is zero no transition to the BC1 ground state takes place.

For negative cubic anisotropy in zero magnetic field the transition from the AF to the 
SF state  takes place at
\begin{equation}
D(F,\Delta)=\frac{1}{2}(1-\Delta)
\end{equation} 
and is  independent of $F$.

\begin{figure} \hspace{-1.5cm} 
  \includegraphics[height=.22\textheight]{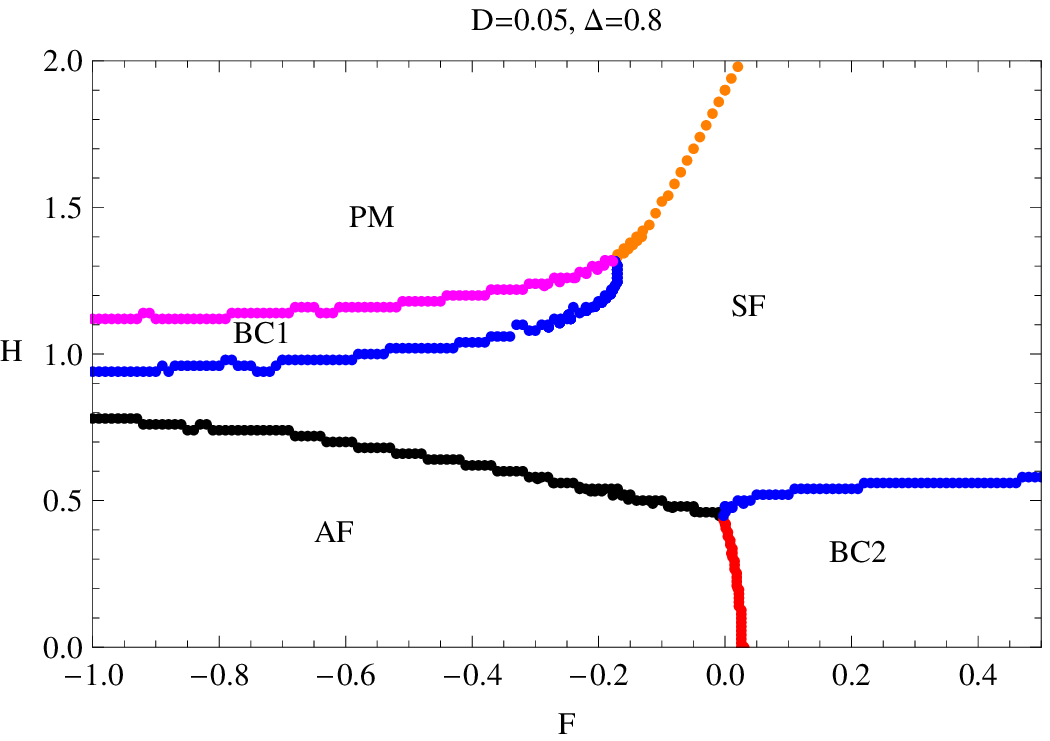}
\includegraphics[height=.25\textheight]{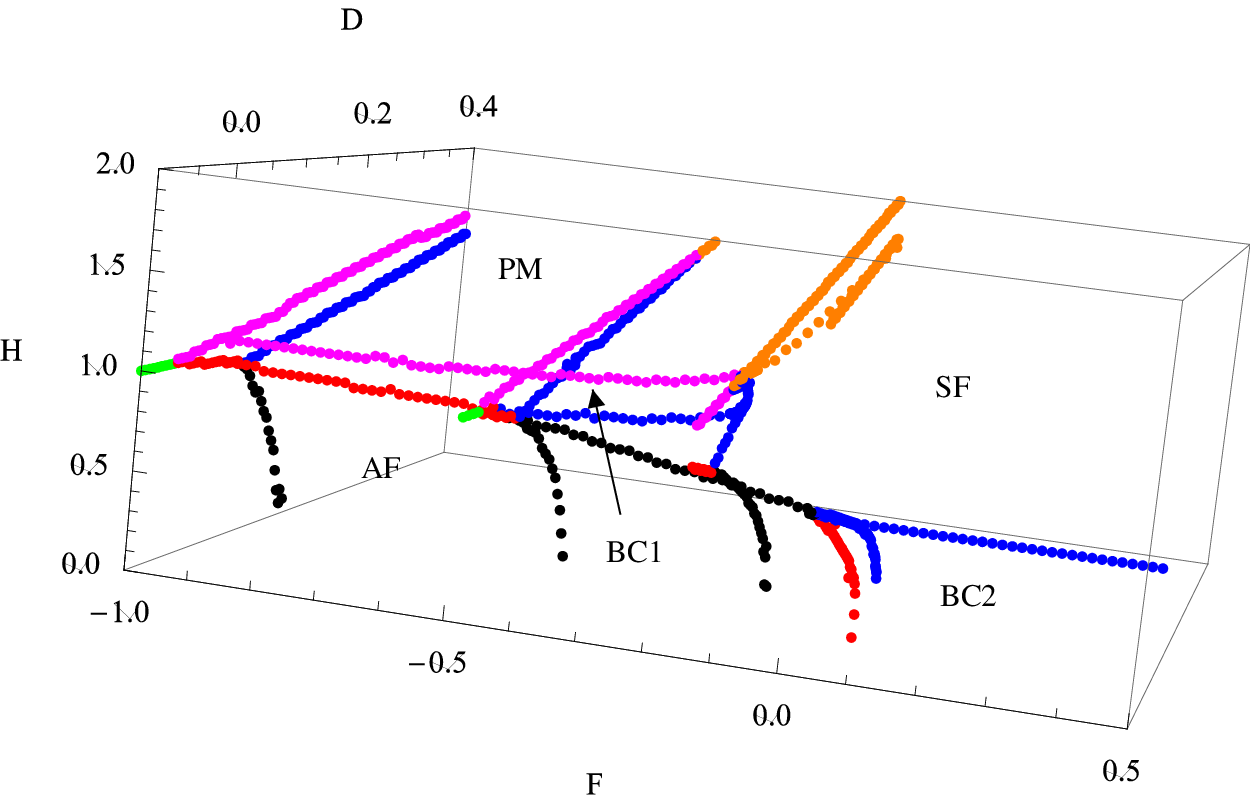} 
  \caption{(online color)Left: Ground states minimizing the energy, Eq. (\ref{all}), in the magnetic field $H$ and
cubic anisotropy $F$ plane at an exchange anisotropy $\Delta=0.8$ and with single ion anisotropy
$D=0.05$\label{all}. Right: A 3D-phase diagram $H(F,D)$ with fix single-ion-anisotropy $\Delta=0.8$}
\end{figure}

\section{Conclusion and outlook}

A system of an anisotropic antiferromagnet shows lots of interesting ground states and
critical points. It can also mapped into a quantum lattice system where the 
biconical ground state is interpreted as the supersolid state. 
We have shown that for the  simple case where just the exchange
($\Delta$) and the single-ion anisotropy ($D$) are considered one can calculate 
analytically all the ground states. In the case with cubic anisotropy 
the resulting Hamiltonian is more difficult to be solved,
 so that we have to calculate the ground states
numerically. In this system two kind of biconical ground states appear, the 
BC1 for $F<0$ and BC2 for $F>0$.

Most of the transition lines are of 
first order, but in some cases the order of the transition also 
changes along the transition line. So lots of interesting critical points
appear ( critical endpoint, tricritical point).  E.g along the 
transition line between BC1 and SF for $F<0$ a reentrance region can be 
shown where the order of the transition changes from second to first 
order by increasing the absolute value of the cubic anisotropy.

However in the absence of the field $H$ we can derive the 
transition surfaces analytically, because the 
term with the applied field ($H$) is the part of the Hamiltonion
which breaks the symmetry  and therefore causes this Hamiltonian
unsolvable. The BC2 ground state does exist also at $H=0$ for
large value of $F$.

The topology of the ground state diagram at $T=0$ determines the possible multicritical points at finite
temperatures. Without cubic anisotropy bicritical or tetracritical points are possible where two lines
of second order phase transitions with order parameters of dimension  $n=1$ and $n=2$ meet.
Cubic anisotropy may change this picture and  multicritical point where two lines with $n=1$ and $n=3$
or $n=3$ and $n=2$ might meet (see Figs. \ref{cub1},\ref{all} for positive values of $F$).
Cubic anisotropy changes most of the transitions to first order, however on the transition line
separating the BC1 from the SF ground state part of this line starting at a critical 
end point is of second order type. Moreover part of the  transition line
separating the spinflop state  from the paramagnetic state is of second order.


\begin{acknowledgments}
  We thank W. Selke for valuable discussions. This work was supported by the Fonds zur F\"orderung der
wissenschaftlichen Forschung under Project No. P19583-N20.  
\end{acknowledgments}

\end{document}